\documentclass[aps,pre,twocolumn,floatfix]{revtex4}

\usepackage{graphicx}
\usepackage{color}
\usepackage[utf8]{inputenc}

\newcommand{\disc}[1]{\hat{#1}}
\newcommand{\ds}{{d_\mathrm{s}}}
\newcommand{\dst}[1]{{d_\mathrm{s}^{(#1)}}}
\newcommand{\sdist}{{\mathcal{D}}}
\newcommand{\nc}{n_\mathrm{c}}
\newcommand{\cm}{\texttt{CM}}
\newcommand{\co}{\texttt{Co}}
\newcommand{\cl}{\texttt{Cl}}
\newcommand{\cocl}{\texttt{Co-Cl}}

\begin{document}

\title{Reconstruction of evolved dynamic networks from degree correlations}
\author{Steffen Karalus}
\email[E-mail: ]{karalus@thp.uni-koeln.de}
\affiliation{Institut f\"ur Theoretische Physik, Universit\"at zu K\"oln,
  Z\"ulpicher Stra{\ss}e 77, D-50937 K\"oln, Germany}
\author{Joachim Krug}
\email[E-mail: ]{krug@thp.uni-koeln.de}
\affiliation{Institut f\"ur Theoretische Physik, Universit\"at zu K\"oln,
  Z\"ulpicher Stra{\ss}e 77, D-50937 K\"oln, Germany}

\begin{abstract}
  We study the importance of local structural properties in networks which have been
  evolved for a power-law scaling in their Laplacian spectrum.  To this end, the
  degree distribution, two-point degree correlations, and degree-dependent clustering
  are extracted from the evolved networks and used to construct random networks
  with the prescribed distributions.  In the analysis of these
  reconstructed networks it turns out that the degree distribution alone is not
  sufficient to generate the spectral scaling and the degree-dependent
  clustering has only an indirect influence.  The two-point correlations are
  found to be the dominant characteristic for the power-law scaling over a
  broader eigenvalue range.
\end{abstract}

\maketitle

\section{Introduction}
\label{sec:intro}
In the mathematical modeling of complex systems, networks have become a
fundamental concept for describing interaction patterns between the constituent
subsystems~\cite{albert_statistical_2002, dorogovtsev_evolution_2002,
  amaral_complex_2004,
  newman_networks:_2010}.  
A major challenge in this field is the relation between network structure and
dynamics: Given the local rules of a dynamical process, how does the interaction
structure shape the overall dynamical behavior?  Although it has been studied
for some time now this question continues to elude a comprehensive answer.
An important bridge between network structure and dynamics, however, was
identified in the spectral properties of a network~\cite{atay_network_2006,
  almendral_dynamical_2007, samukhin_laplacian_2008,vanMieghem2011,Jalan2015,Grabow2015}.  The eigenvalues (and
eigenvectors) of network matrices are known to encode global structural
properties as well as the overall dynamical behavior.  The probably most
prominent example in this context is the graph Laplacian.  
Besides very important structural properties such as the algebraic
connectivity~\cite{chung_spectral_1997}, the Laplacian spectrum determines the
overall behavior of fundamental processes such as
synchronization~\cite{arenas_synchronization_2008}, vibrational modes of
Gaussian polymer structures~\cite{gurtovenko_generalized_2005}, and random walks
or diffusion~\cite{havlin_diffusion_2002, noh_random_2004}.

A second dynamical aspect is the evolution of network structure.  In many
systems, the internal connectivity may change with time.  If both processes,
dynamics on and evolution of the network, are present the relation between their time
scales becomes important.  In the case of a separation of time scales with fast
dynamics and slow evolution, it is usually the overall behavior of the dynamics
which guides the evolution.  This principle has been adopted as an optimization
strategy and applied in many different contexts including Boolean threshold
dynamics~\cite{bornholdt_topological_2000, oikonomou_effects_2006}, the
emergence of modularity in changing
environments~\cite{kashtan_spontaneous_2005}, the synchronizability of
oscillator networks~\cite{donetti_entangled_2005, rad_efficient_2008}, and
reconstruction of networks from their Laplacian
spectra~\cite{ipsen_evolutionary_2002, comellas_spectral_2008}.

In the investigation of networked systems, random network models have always
played a central role.  They serve as null models that are conditioned to satisfy certain prescribed 
properties while being 
maximally random in every other sense.  Random network models
range from the very simplistic Erd\H{o}s-R\'{e}nyi random
graphs~\cite{erdos_random_1959} to sophisticated models able to reproduce many
nontrivial network properties~\cite{newman_equitable_2014}.  By analyzing such
random networks, the relevance of the prescribed structural properties in
empirical or artificially generated networks can be tested.

\begin{figure}
  \centering
  \includegraphics[width=\linewidth]{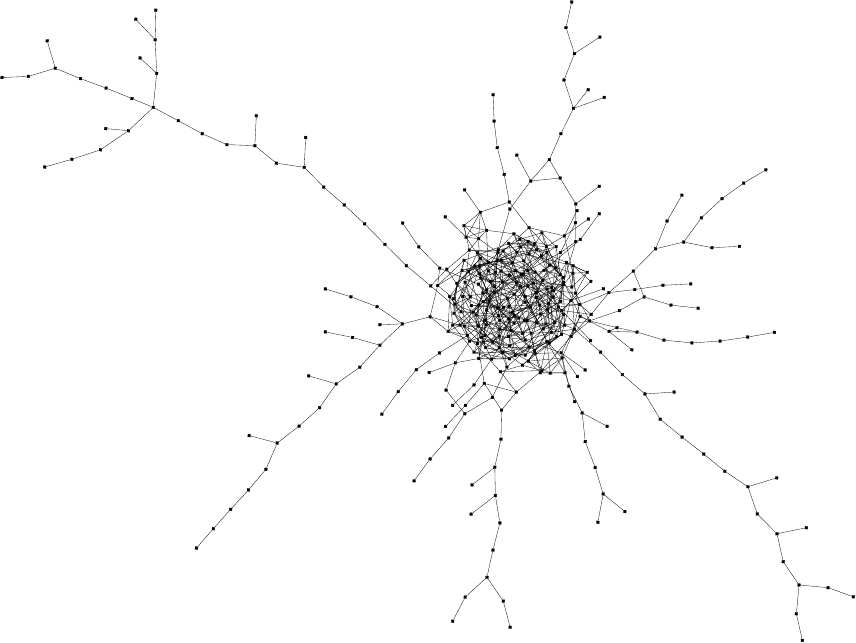}
  \caption{Typical realization of a network with $N=361$ vertices and $M=722$
    edges evolved towards a power law Laplacian spectrum with spectral
    dimension $\dst{1}=1.4$ after $10^6$ evolution steps.}
  \label{fig:net-evolved}
\end{figure}

In a preceding study~\cite{karalus_network_2012}, networks were successfully
evolved towards an approximative power-law scaling of the integrated
spectral density $I(\lambda) \propto \lambda^{\ds/2}$ of the graph Laplacian with a prescribed non-trivial
exponent, the so-called spectral dimension~$\ds$.  These networks generate
anomalous diffusion behavior described by a power-law decay with the same
exponent in the average return probability of a random walker $P_0(t) \propto
t^{-\ds/2}$. A summary of the mathematical approach underlying 
the evolution is presented in the Appendix. The degree
distributions of the evolving networks as well as correlations, measured by the
assortativity and clustering coefficients, were observed to change significantly
in the course of the evolution.  The emergence of a bimodal degree distribution
together with increasing degree assortativity and clustering are strong
indications for heterogeneous network structures evolving out of the homogeneous
initial configurations, two-dimensional square lattices and connected
(Erd\H{o}s-R\'{e}nyi) random graphs of the same size.  The structures of the
evolved networks are mainly characterized by the presence of two distinguished
regions, densely connected cores of high-degree vertices on the one hand and
sparsely connected peripheries of low-degree vertices on the other hand.  Such
core-periphery structures can indeed be identified in the exemplary evolved
network configuration depicted in Fig.~\ref{fig:net-evolved}.

To which extent do these structural properties determine the spectral and,
consequently, the dynamical behavior of the networks?  Is there a way to
construct random networks with power-law Laplacian spectra from scratch based on
the correlation functions?  These questions are addressed in the following.  The
correlations are extracted from the evolved networks and used to generate random
networks following the prescribed distributions.

\section{Definitions and Algorithms}
\label{sec:definitions}
The formal description of a network with $N$ vertices and $M$ edges is usually
given by the $N \times N$ adjacency matrix $A$.  Its elements are $A_{ij} = 1$
if vertices $i$ and $j$ are connected by an edge and $A_{ij} = 0$ otherwise.
Any simple network (undirected with neither multiedges nor self-loops) can be
equivalently described by the graph Laplacian $L = D - A$ where $D$ is the
diagonal matrix of vertex degrees, $D_{ij} = k_i \delta_{ij}$ (with
$\delta_{ij}$
being Kronecker's delta).  The vertex degree $k_i = \sum_j A_{ij}$ is the basic
structural property of a given vertex $i$ in a network, counting its connections
to other vertices.

A systematic statistical description of network properties 
starts out from the distribution of vertex degrees and then expands
into two-point, three-point, etc.\ correlations between vertex degrees.  The
degree distribution $P(k)$ is the probability that a randomly chosen vertex has
degree $k$.  Its discrete counterpart $\disc{P}(k)$ denotes the number of
vertices with degree $k$ in a given network.  Two-point correlations are
described by the joint degree distribution $P(j,k)$, the probability that a
randomly chosen edge connects vertices of degrees $j$ and $k$.  Its discrete
counterpart $\disc{P}(j,k)$ denotes the number of edges between vertices with
degrees $j$ and $k$ for a given network.  The overall degree-degree correlations
are quantified by the assortativity
coefficient~$r$~\cite{newman_assortative_2002, newman_mixing_2003}.  It is the
Pearson correlation coefficient of the degrees of adjacent vertices with values
ranging from $-1$ to $1$. The value $r=0$ means that the degrees of neighboring vertices
are uncorrelated, whereas $r>0$ ($r<0$) indicates positive (negative)
correlations and the network is said to be (dis)assortative.  A full systematic
treatment of three-point correlations is rather involved.  Instead, these are
usually subsumed in clustering coefficients.  The (global) clustering
coefficient~$C$, defined as the density of triangles in a
network~\cite{newman_random_2001}, measures the overall transitivity, i.e., the
tendency for two neighbors of the same vertex also to be neighbors of one
another.  This can also be defined for each vertex individually, yielding the
local clustering coefficient of a vertex~$i$, defined as $C_i = 2 T_i / k_i
(k_i-1)$~\cite{watts_collective_1998}. Here $T_i$ denotes the number of edges
between neighbors of $i$ which is also the number of triangles in which vertex
$i$ takes part.  Averaging the local clustering coefficient over all vertices in the same
degree class yields the degree-dependent
clustering~$C(k)$~\cite{vazquez_large-scale_2002}.  All clustering coefficients
range from $0$ (no transitivity, i.e., no triangles present) to $1$ (complete
transitivity, i.e., all components are complete subgraphs).  When working with
absolute frequencies, the degree-dependent clustering can be expressed by
$\disc{T}(k) = C(k) \disc{P}(k) k (k-1) / 2$, the number of triangle corner
vertices with degree $k$.

\subsection*{Random networks with prescribed correlations}
Several algorithms have been proposed to construct random networks with a given
degree distribution and correlations between vertex degrees.  Mostly, these are
extensions of the configuration model algorithm~\cite{molloy_critical_1995}, in
the following abbreviated by \cm{}.  The basic idea of the configuration model
is to first assign a number of half-edges according to a discrete degree
sequence specified by $\disc{P}(k)$ to each vertex and then randomly pair these
half-edges in order to form the edges of the networks.  If the algorithm
succeeds, the result is a random network with exactly the chosen degree
sequence.  

Ángeles Serrano and Boguñá~\cite{angeles_serrano_tuning_2005} extended the
configuration model algorithm to incorporate the degree-dependent
clustering~$C(k)$ as additional specification of the random networks to be
constructed.  In a two-step process, first those half-edges are selected to be
paired that form additional triangles in the classes of vertex degrees in which
the prescribed $\disc{T}(k)$ has not yet been reached.  Secondly, remaining free
half-edges are matched randomly.  Throughout this paper, this algorithm is
denoted by \cl{}.  

A different approach was proposed by Weber and
Porto~\cite{weber_generation_2007} for the generation of networks with
prescribed degree-degree correlations $P(j,k)$ specified by $\disc{P}(j,k)$
which implicitly defines the degree distribution $P(k)$ as well.  In this
algorithm, here denoted by \co{}, the half-edges to be paired are selected
according to the remaining number of edges to be built between vertices of the
corresponding degree class.  

In order to incorporate both, degree-degree correlations and clustering, into
random network generation, Pusch et al.~\cite{pusch_generating_2008} extended
the \cl{}-algorithm to additionally include prescribed two-point correlations
$P(j,k)$.  In this algorithm, denoted here by \cocl{} , all three absolute
frequencies $\disc{P}(k)$, $\disc{P}(j,k)$, and $\disc{T}(k)$ have to be
specified as input.  The network generation consists of three steps.  First,
half-edges are selected by the degree of the vertices to which they are
connected under the conditions that in the corresponding degree classes the
prescribed number of edges has not been reached and that their pairing forms
triangles in the degree classes that have not reached the prescribed number of
triangle corners.  Secondly, random half-edges are chosen to form edges in the
degree classes which are not yet satisfied.  Thirdly, still remaining half-edges
are paired randomly as in the configuration model.  The algorithm will, if
possible, exactly reproduce the prescribed degree sequence, very closely
resemble the given two-point correlations, and approximately generate the
prescribed numbers of triangle corners.

Here, the \cocl{}-algorithm will be used for the ``full'' reconstruction of the
evolved networks in Sec.~\ref{sec:reconstr-full}.  In
Sec.~\ref{sec:reconstr-less}, the other three algorithms \cm{}, \cl{}, and \co{}
are employed individually in order to determine which of the correlations is most relevant
for the replication of the power-law spectrum.  Subsequently, in Sec.~\ref{sec:reconstr-ts} 
not only the final state of the network evolution is used but the correlations are extracted for
all intermediate states of a single network evolution trajectory and fed into all four
random network generation algorithms. This
provides further insight into the relevance of the correlation measures at
different stages of the evolution.

\section{Reconstruction using all correlations}
\label{sec:reconstr-full}
Different realizations of the evolved networks are reconstructed individually by
the \cocl{} algorithm.  The network configurations are taken from the evolution
of random networks towards the target spectral dimension $\dst{1} = 1.4$ as
presented in Ref.~\cite{karalus_network_2012}.  The degree distribution, the
degree-degree correlations, and the degree-dependent clustering of the final
configurations, i.e., after $10^6$ evolution steps, are calculated as absolute
frequencies $\disc{P}(k)$, $\disc{P}(j,k)$, and $\disc{T}(k)$.  This is done
individually for 100 realizations of the evolution.  For each of the triples
$\disc{P}(k)$, $\disc{P}(j,k)$, and $\disc{T}(k)$, 100 samples of the random
networks are generated and analyzed.

\begin{figure}
  \centering
  \includegraphics[width=\linewidth]{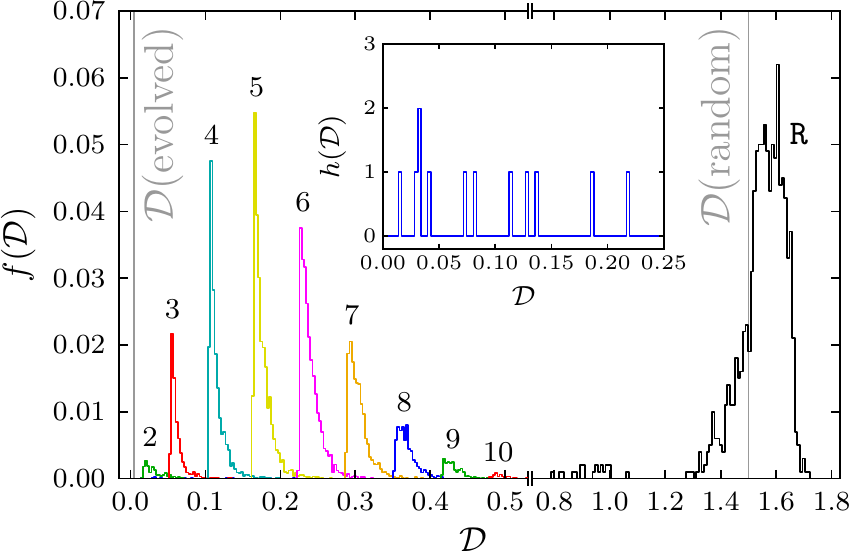}
  \caption{(Color online) Histogram of spectral distances in the reconstructed
    networks.  Shown is the fraction $f(\sdist)$ of networks with spectral
    distance $\sdist$ to the evolution target reconstructed from the evolved
    networks (colored lines with number labels) and a random network of the same
    size (black line, labeled \texttt{R}).  In the former, the different colors
    stand for different numbers of connected components~$\nc$, indicated by the
    numbers above the curves. Networks with $\nc=1$ and $\nc=11$ appear with
    very low frequencies.  The (average) spectral distances of the original evolved and random networks
    are indicated by the vertical gray lines.  The inset shows an enlarged segment 
    of the histogram for $\nc=1$, now in terms of absolute frequencies
    $h(\sdist)=f(\sdist)\times 10^4$.}
  \label{fig:distance_hist_full}
\end{figure}
In order to see how well the reconstructed networks reproduce the evolution
target, i.e., the prescribed power-law scaling in the Laplacian spectrum, the
first quantity to look at is the spectral distance~$\sdist$
to the evolution target (defined in Eq.~(\ref{eq:sdist}) of the Appendix).  The
value of $\sdist$ is lower the closer the Laplacian spectrum resembles the
prescribed power law given by the value of $\ds$.
Figure~\ref{fig:distance_hist_full} displays the distribution~$f(\sdist)$ of the
spectral distances for the reconstructed networks.  For comparison, the
distributions $\disc{P}(k)$, $\disc{P}(j,k)$, and $\disc{T}(k)$ were also
calculated for an uncorrelated random network of the same size, the initial
configuration of the evolutionary optimization.  These distributions were used
for the generation of 1000 realizations of reconstructed random networks by the
same algorithm.  The spectral distance of the reconstructed evolved networks is
always higher than the average value of the evolved networks but significantly
lower than the values of the random network and its reconstructed networks.
Hence, the degree distribution and the two correlation measures together indeed
encode the spectral behavior to a significant extent. 

Remarkably, the
distribution shows a structure of several, mostly well-separated peaks.  As
indicated by the different colors in Fig.~\ref{fig:distance_hist_full}, the
multi-peak structure is generated by different numbers of connected
components~$\nc$ in the reconstructed networks. Recall that 
a connected component of an undirected network is a
subgraph in which any two vertices are connected to each other by
paths, but are disconnected from all vertices in the other
components. In the present context this means that a random walker
diffusing on the network is confined to the connected component in
which it started, and hence the diffusion processes on different
components are independent. In contrast to the network
evolution, which is explicitly constrained to produce a globally
connected graph, the construction of the correlated random networks provides no
possibility to control the number of components. 
 Only a very small fraction of the resulting reconstructed networks,
namely $12$ out of $10^4$, are found to be globally connected like the evolved
networks. An explanation for the separation of the distribution into classes of
equal number of connected components is that the number of components dictates
the degeneracy of the smallest eigenvalue $\lambda_1 = 0$ and the smallest
eigenvalues have the largest influence on the spectral distance.
Notably, the
globally connected networks with $\nc=1$ deviate from the trend and do not
exhibit spectral distances lower than those with $\nc=2$.  Due to the small
number of realizations this observation can, however, not be considered as
statistically significant.

\begin{figure}
  \centering
  \includegraphics[width=\linewidth]{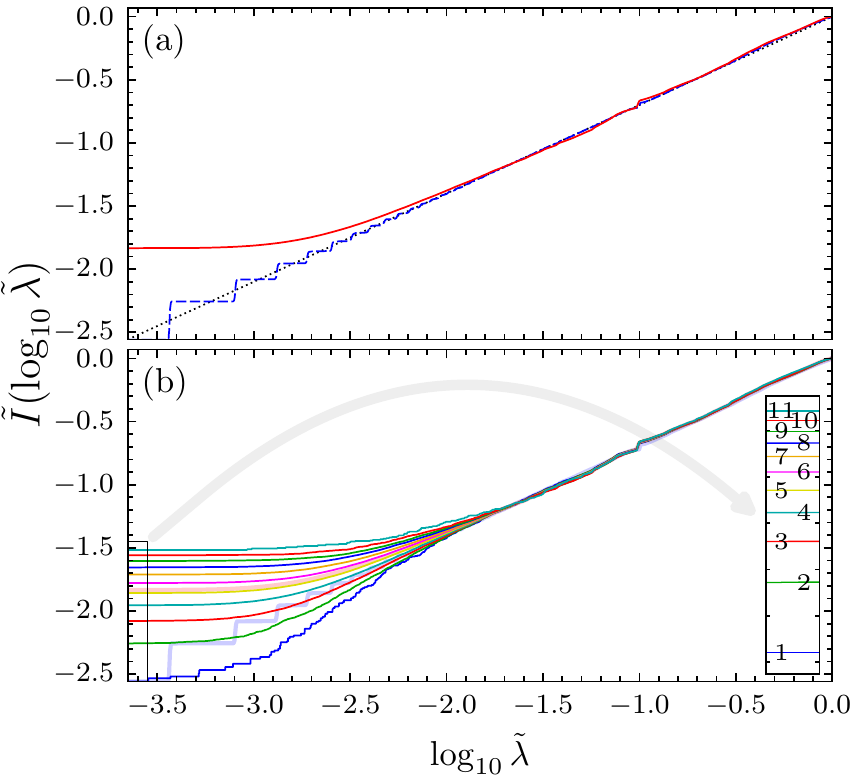}
  \caption{(Color online) Averaged logarithmically integrated Laplacian spectral
    density of reconstructed networks.  In the upper panel~(a), the averaged
    spectral densities of the evolved (dashed blue line) and the reconstructed
    (red line) networks are shown.  The black dotted line displays the evolution
    target.  The latter is broken down according to the number of connected components in the
    lower panel~(b).  The curves are labeled by their numbers of components in
    the inset, the color code is the same as in
    Fig.~\ref{fig:distance_hist_full}.  For comparison, the transparent lines in
    the background show the curves of panel (a) again.}
  \label{fig:density_full}
\end{figure}
The influence of the small eigenvalues on the spectral distance is also clearly
visible in the integrated spectral densities.  Figure~\ref{fig:density_full}(a)
displays the averaged logarithmically integrated spectral densities (defined in
Eq.~(\ref{eq:logint_ev_dens}) of the Appendix)
of the evolved and reconstructed networks.  Evidently, the higher frequency
of small eigenvalues is the main cause for the larger deviation from the target
function.  In Fig.~\ref{fig:density_full}(b) the spectral densities are
individually averaged for the different classes of equal number of connected
components.  It shows how the deviation from the target function in the region
of small eigenvalues indeed increases with the number of connected components in
the network.  The increasing degeneracy of the zero eigenvalue makes the
integrated densities start out from higher values so that an increasing initial
exceedance of the target function is inevitable.  The integrated densities of
the globally connected networks with $\nc = 1$ actually fall below the target
function explaining the higher spectral distance than for $\nc = 2$ although the
global trend of a lower initial value is continued.  But again, due to the small
number of realizations this observation should not be considered as
statistically significant.

The shape of the spectral densities shown in
Fig.~\ref{fig:density_full}(b) suggests a simple approximation that
accounts for the positions of the peaks in the histogram of spectral
distances in Fig.~\ref{fig:distance_hist_full}. Assuming that the
spectral density exactly follows the target power law for large
eigenvalues and becomes constant once the degeneracy $n_c$ of the
minimal eigenvalue is reached, the evaluation of the spectral
distance measure (\ref{eq:sdist}) yields the relation $\sdist(n_c) =
\frac{2}{3 d_s} \vert \log(n_c) \vert^3$, which is in good agreement
with the numerical data for $n_c \geq 2$.

\section{Reconstruction using partial information}
\label{sec:reconstr-less}
Having seen that reconstructing the power-law Laplacian spectrum by using the
degree distributions, degree-degree correlations, and degree-dependent
clustering simultaneously works reasonably well, the question arises which
out of these three measures is most relevant.  In order to tackle this question,
the distributions $\disc{P}(k)$, $\disc{P}(j,k)$, and $\disc{T}(k)$ extracted
from the evolved networks are used as input for the four algorithms \cm{},
\cl{}, \co{}, and \cocl{}.  Again, this is done independently for $100$
realizations of the evolution and for each realization $100$ samples of random
networks are generated.

\begin{figure}
  \centering
  \includegraphics[width=\linewidth]{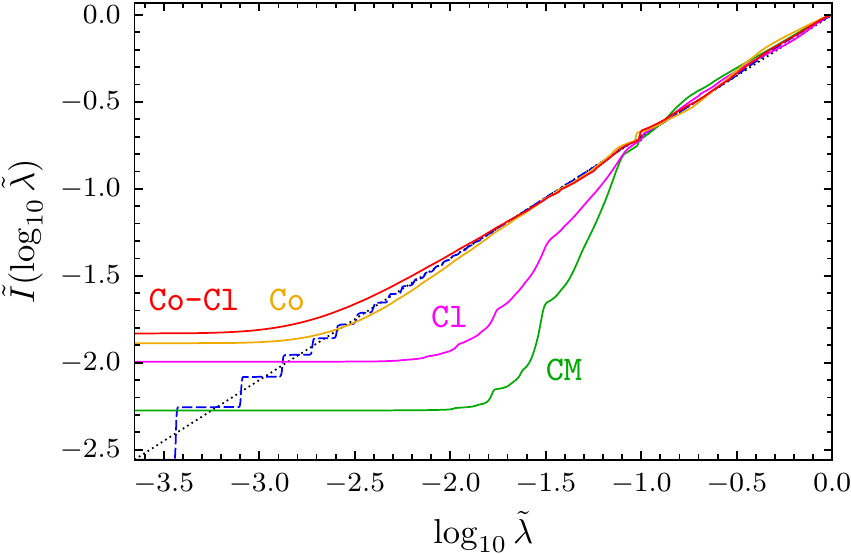}
  \caption{(Color online) Averaged logarithmically integrated Laplacian spectral
    densities of evolved networks (dashed blue) and reconstructions by the \cm{}
    (green), \cl{} (magenta), \co{} (orange), and \cocl{} (red) algorithms.  The
    black dotted line displays the evolution target.}
  \label{fig:density_all}
\end{figure}
Figure~\ref{fig:density_all} shows the logarithmically integrated Laplacian
spectral densities averaged over all $10^4$ samples for the four algorithms.
First of all, we observe that for eigenvalues larger than a certain value
(around $\log_{10} \tilde\lambda = -1$ which turns out to be the eigenvalue $\lambda
= 1$ \footnote{The maximum eigenvalue is found to be approximately
  $\lambda_{\max} \approx 10$ in all cases.  Hence, $\log_{10} \tilde\lambda
  \approx -1$ corresponds to $\lambda \approx 1$.}) all reconstruction
algorithms reproduce the power law spectrum fairly well.  Hence, the
specification of the degree distribution appears to be sufficient for this part
of the Laplacian spectrum.  For smaller eigenvalues, the reconstructions by the
\cm{} and \cl{} algorithms, i.e., those algorithms not using the degree-degree
correlations, deviate substantially.  Networks constructed by the \cm{}
algorithm show the largest deviation from the power law while those generated by
\cl{} lie slightly closer but differ in the same range of
eigenvalues. In both cases the integrated spectral
densities display several kink-like features which reflect 
accumulations of eigenvalues. A possible mechanism underlying such
spectral degeneracies are network symmetries
\cite{karalus_symmetry-based_2015}. These features are absent from the spectra 
obtained using the \co{} and \cocl{} algorithms, which 
both resemble the target power law closely and almost equally well. Thus, the
degree-degree correlations seem to be the major structural factor causing the
scaling.  The degree-dependent clustering has only a minor influence when the
two-point correlations are reproduced as well.

\begin{figure}
  \centering
  \includegraphics[width=\linewidth]{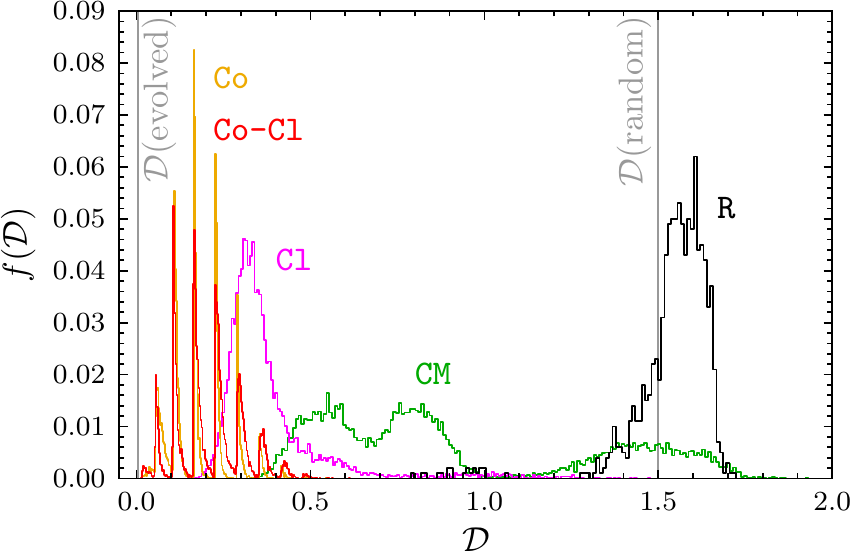}
  \caption{(Color online) Histogram of spectral distances for the 
    networks reconstructed by all four algorithms.  Shown is the fraction $f(\sdist)$ of
    networks with spectral distance $\sdist$ to the evolution target
    reconstructed from the evolved networks by the \cm{} (green), \cl{}
    (magenta), \co{} (orange), and \cocl{} (red) algorithms.  The black
    histogram (\texttt{R}) is the same as in Fig.~\ref{fig:distance_hist_full}
    and the (average) spectral distances of the original evolved and random 
    networks are indicated by the vertical gray lines.}
  \label{fig:distance_hist_all}
\end{figure}
For a quantification of this observation, Fig.~\ref{fig:distance_hist_all}
displays histograms of the spectral distances for all four reconstruction
algorithms.  As expected, the networks reconstructed by the \co{} and \cocl{}
algorithms have the lowest spectral distances to the power-law spectrum.  Both
histograms are very similar and exhibit the characteristic multi-peak structure
seen in Fig.~\ref{fig:distance_hist_full}.  The spectral distances of the
networks generated by \cl{} and \cm{} are significantly larger, but even for the
latter ones on average still lower than those of the reconstructed random
networks.

One approach to avoid the inconveniences of handling not globally connected
networks, such as the observed strong dependency of the spectral distance on the
number of connected components, is to restrict the analysis to the largest
components of the generated networks.  By doing so, one presumes that the
largest component represents the network as a whole and smaller components are
less important.  A major problem arises in the comparison of these restricted
networks which will naturally all have different sizes.  As the spectral
distance depends inherently on the number of vertices~$N$ there is no way to
compare its values for networks of different sizes.
\begin{figure}
  \centering
  \includegraphics[width=\linewidth]{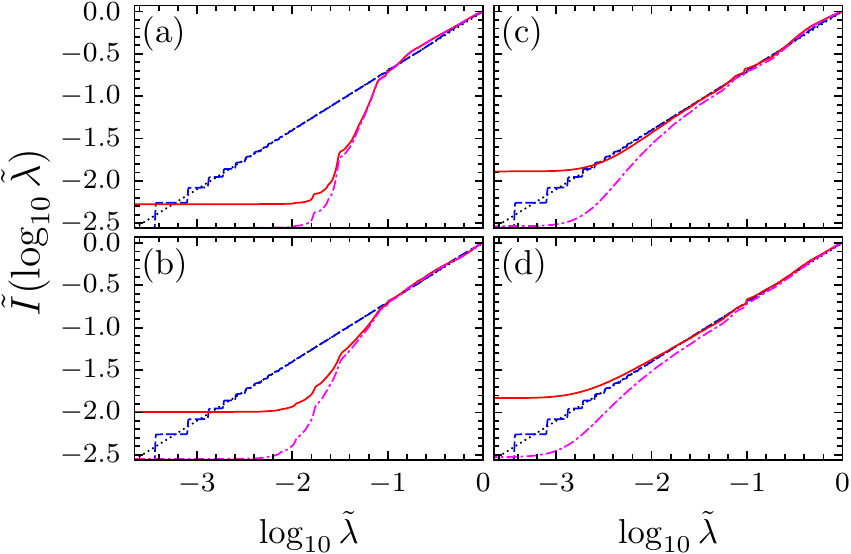}
  \caption{(Color online) Averaged logarithmically integrated Laplacian spectral
    densities of evolved networks (dashed blue), networks reconstructed by the
    (a)~\cm{}, (b)~\cl{}, (c)~\co{}, and (d)~\cocl{} algorithms (red) as well
    as their largest components (dash-dotted magenta).}
  \label{fig:density_lc_all}
\end{figure}
Nevertheless, the spectra can be calculated and visually compared.  This is done
in Fig.~\ref{fig:density_lc_all}.  The first observation is that the degeneracy
of the smallest eigenvalue $\lambda = 0$ is removed in the spectra of the
largest components.  In all four cases, the averaged logarithmically integrated
spectral densities of the full networks and the respective largest components
are qualitatively comparable.  The deviations affect mainly the region of low
eigenvalues where the integrated densities are significantly lower for the
isolated largest components.  In the \cm{} and \cl{} reconstructions this
results in a further deviation from the target power law.  For the \co{} and
\cocl{} reconstructions it might be difficult to tell if the spectra of the full
networks or the isolated largest components lie ``closer'' to the target.
However, the full networks resemble the power law over a larger range of
eigenvalues than their largest components.

\section{Reconstruction of time series}
\label{sec:reconstr-ts}
In this section, the relevance of the different correlation measures in the
evolutionary process shall be examined.  To see if the degree distribution,
degree-degree correlations, and degree-dependent clustering equally well
characterize the network configurations at different stages of the evolution
process, the reconstruction is applied to all intermediate steps of one
exemplary evolution from a random graph towards the target spectral dimension of
$\dst{1}=1.4$.  As before, the distributions are extracted as absolute
frequencies $\disc{P}(k)$, $\disc{P}(j,k)$, and $\disc{T}(k)$, now for each
evolutionary time step.  For each point in this time series, $100$ samples of
the random (correlated) networks are generated independently by all four
algorithms and analyzed.

\begin{figure}
  \centering
  \includegraphics[width=\linewidth]{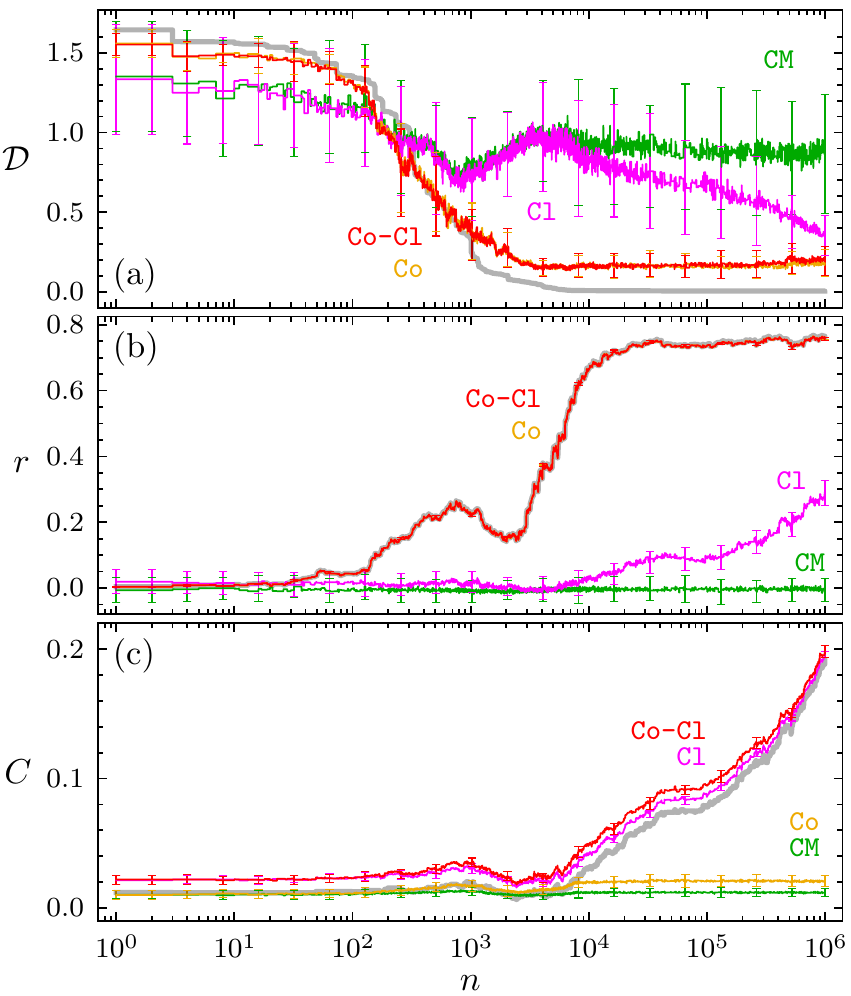}
  \caption{(Color online) Reconstruction averages of evolutionary time series by
    the \cm{} (green), \cl{} (magenta), \co{} (orange), and \cocl{} (red)
    algorithms.  Shown are (a)~the spectral distance~$\sdist$ to the evolution
    target, (b)~the assortativity coefficient~$r$, and (c)~the global clustering
    coefficient~$C$.  All averages are calculated over 100 samples of the
    reconstruction and the error bars mark one standard deviation.  The light
    gray curves in the background display the time series of the respective
    quantities in the evolving network.}
  \label{fig:ts_all}
\end{figure}
The results of the time series reconstruction are summarized in
Fig.~\ref{fig:ts_all} showing (a)~the spectral distances~$\sdist$, (b)~the
assortativity coefficients~$r$, and (c)~the global clustering coefficients~$C$
of the reconstructed networks by all four algorithms at each time step~$n$ of
the evolution in comparison with the respective values from the network
evolution.  The spectral distances of all reconstructed networks are observed to
take values in the same range as the evolving network for approximately the
first $10^2$ evolution steps.  For the reconstructions including the
degree-degree correlations, i.e., the \co{} and \cocl{} algorithms, this trend
continues for roughly another decade.  Afterwards, from around $10^3$ evolution
steps on, all reconstructed networks have larger spectral distances to the
target power law than the evolving network itself with the \co{} and \cocl{} algorithms
reaching significantly smaller values than \cm{} and \cl{}.  Additionally, after
around $10^4$ evolution steps the spectral distance of the networks generated by
\cl{} is observed to decrease.  The reconstructions including the degree-degree
correlations as input, \co{} and \cocl{}, exactly reproduce the assortativity
coefficient of the evolving network, which increases and saturates to a rather high value after around $10^4$
evolutionary steps, throughout the time evolution.  
The \cm{} algorithm generates non-assortative networks during the whole
evolution while the networks constructed by the \cl{} algorithm show an
increasing assortativity simultaneously with the observed decrease in the
spectral distance in these networks after approximately $10^4$ evolutionary
steps.  The reconstructions including the degree-dependent clustering, \cl{} and
\cocl{}, reproduce the increasing clustering coefficient of the evolution while
the \co{} and \cm{} algorithms generate networks with no transitivity.

These observations can be interpreted in the following way.  In the initial
phase of the evolution up to around $10^2$ evolution steps, slight improvements
towards the power-law spectral densities are governed by the degree
distribution.  All reconstructions, including the \cm{} algorithm, follow this
trend.  In the second phase up to around $10^3$ evolution steps, changes in the
two-point correlations towards assortative structures are prevalent.  These
result in the largest reduction of the spectral distance.  As only the \co{} and
\cocl{} algorithms are able to reproduce this change they follow the reduction
in the spectral distance.  For the remaining evolution steps, more refined
structural changes take place in the network evolution.  These are not described
by the degree distribution and correlation measures so that none of the
algorithms is able to reproduce the improvement in the spectral distance.  They
are, however, accompanied by an increase in transitivity from around $10^4$
evolution steps on which is reproduced by the \cl{} and \cocl{} algorithms.
This does not influence the spectral distance of the networks reconstructed by
the \cocl{} algorithm.  For the reconstruction by the \cl{} algorithm, the
increase in transitivity is accompanied by an increase in assortativity and a
decrease in the spectral distance at the same time.  It is known that clustering
and assortativity are correlated~\cite{angeles_serrano_tuning_2005}, so the
observation is not surprising.  In this way, the networks reconstructed by \cl{}
indirectly acquire a higher assortativity which lets them ``catch up'' with the
\co{} and \cocl{} reconstructions to some extent and also attain lower spectral
distances.

\section{Conclusions}
\label{sec:conclusions}
We studied the importance of the distribution of vertex degrees and their
correlations as basic structural measures in networks evolved towards a
power-law Laplacian spectrum with a prescribed spectral dimension.  To this end,
random networks were generated and analyzed with the same degree distribution,
degree-degree correlations, and degree-dependent clustering as the evolved
networks.  In this reconstruction procedure, the degree-degree correlations
turned out to be the most important measure for the spectral scaling.

\begin{figure*}
  \centering
  \includegraphics[width=.9\linewidth]{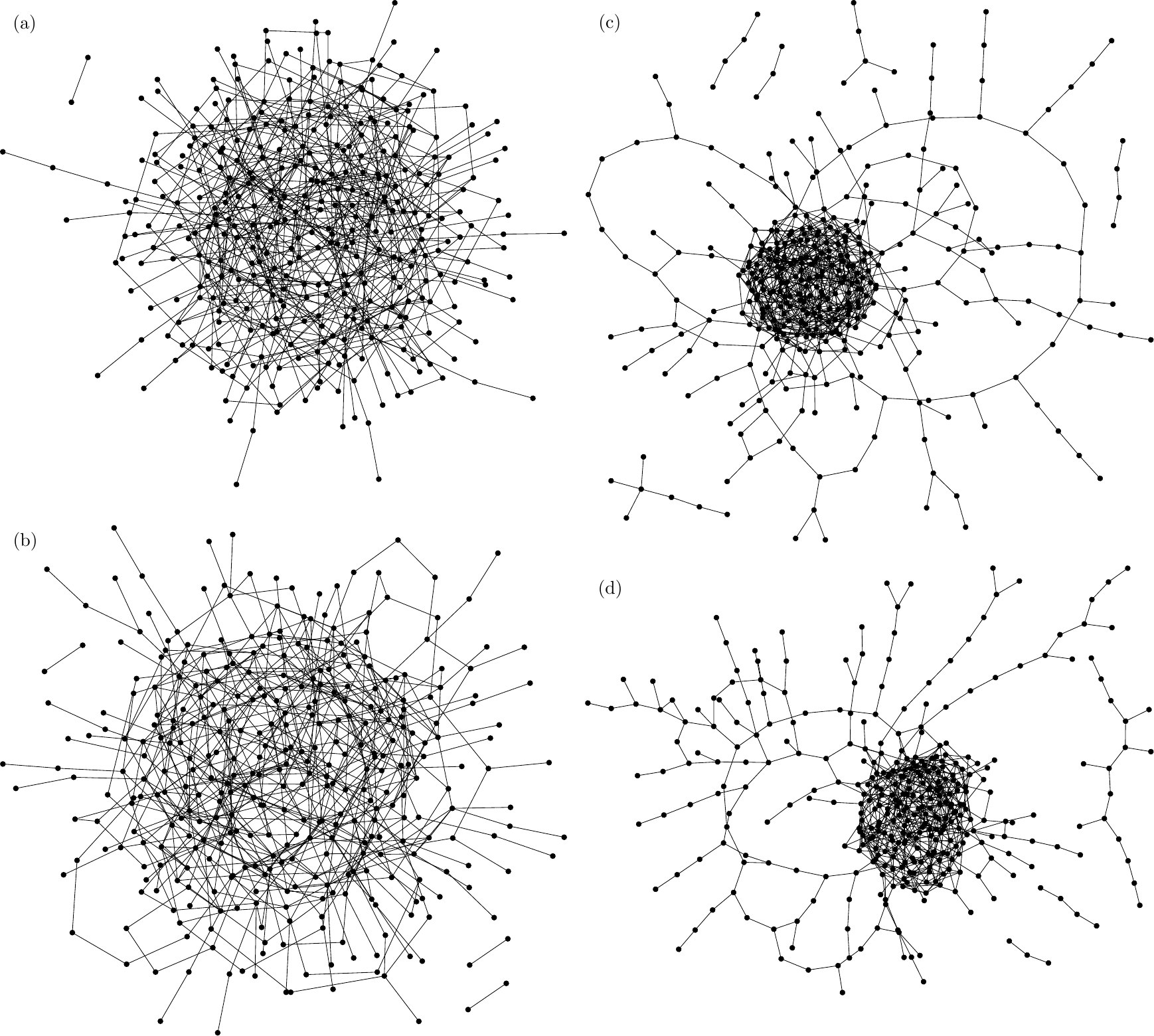}
  \caption{Typical configurations of reconstructed evolved networks making use
    of different correlation measures: (a)~\cm{}-algorithm, degree distribution
    only, (b)~\cl{}-algorithm, degree distribution and degree-dependent
    clustering, (c)~\co{}-algorithm, degree distribution and degree-degree
    correlations, and (d)~\cocl{}-algorithm, degree distribution,
    degree-dependent clustering, and degree-degree correlations.}
  \label{fig:net_reconst_all}
\end{figure*}
The evolved networks were predominantly characterized by a bimodal degree
distribution and a high assortativity, i.e., positive degree-degree
correlations.  Additionally, an increasing transitivity, measured by the
clustering coefficient, was observed.  Together, they indicate a structural
segregation into two distinct regions, densely connected cores and sparse
peripheries.  The results presented here confirm that assortativity is indeed
the main structural feature of the evolved networks and higher order
correlations, as measured by the degree-dependent clustering, play a minor role
only.
This interpretation is enforced by looking at typical realizations of the
reconstructions from all four algorithms in Fig.~\ref{fig:net_reconst_all}: The
\co{}- and \cocl{}-networks show signs of core-periphery structures as seen in
the evolved networks (see Fig.~\ref{fig:net-evolved}) while the \cm{}- and
\cl{}-networks appear rather homogeneous.  Secondly, it was observed that
networks with a high clustering coefficient generated by the \cl{} algorithm
appear to resemble the power-law spectrum better than those without (generated
by the \cm{} algorithm).  This seems to be an indirect effect mediated by the
correlations between assortativity and clustering: Networks with high clustering
are known to be always assortative.

Finally, we remark that at this point we cannot answer the question why the
observed core-periphery structures generate the power-law Laplacian spectra and,
thus, anomalous diffusion behavior.  In a different setting---where evolving
networks were kept homogeneous---it was observed that networks with loops and
dangling ends of different lengths are also able to generate such
behavior~\cite{karalus_symmetry-based_2015}.  A distribution of paths of
different lengths through the network might as well be generated by the
core-periphery structures found here.  Just like in comb-like networks with
power-law distributed teeth lengths~\cite{havlin_diffusion_2002} this
interpretation provides an intuitive view on how the anomalous diffusion
behavior is generated. The development of a minimal,
  analytically tractable model that elucidates the relationship
  between the core-periphery structure and the resulting power law
  spectrum seems highly desirable, but must be left to future work.

\section*{Acknowledgments}
We thank Markus Porto for his support in the design of this study as well as
Sebastian Weber and Andreas Pusch for providing their implementations of the
different algorithms.  We gratefully acknowledge partial funding by the
\emph{Studienstiftung des deutschen Volkes} and the \emph{Bonn-Cologne Graduate
  School of Physics and Astronomy}.

\appendix
\section{Network evolution}
Network evolution as a method to construct network structures with a prescribed
power-law scaling in the Laplacian spectrum was successfully developed and
described in Ref.~\cite{karalus_network_2012}.  The idea is to explore the
configuration space of valid networks by successive steps of mutation and
selection.  In the basic setting, the mutation is realized by a random rewiring
of one edge.  The selection accepts this mutation if a spectral distance
function~$\sdist$ (called $\Delta$ in the original reference) is lowered and the
network remains globally connected.  In the following, the formal concepts of
the network evolution are briefly summarized.

In order to represent the eigenvalue spectrum of the graph Laplacian,
$\{\lambda_\nu\}_{\nu=1,\ldots,N}$, in a functional form the integrated
spectral density
\begin{equation}
  \label{eq:int_ev_dens}
  I(\lambda) = \frac{1}{N} \sum_{\nu=1}^{N} \Theta(\lambda - \lambda_\nu)
\end{equation}
is a convenient choice.  Here, $\Theta$ denotes the Heaviside step function,
$\Theta(x) = 1$ for $x \geq 0$ and $\Theta(x) = 0$ for $x < 0$.  Rescaling by
the maximum eigenvalue, $\tilde\lambda_\nu = \lambda_\nu /
\max_{\nu^\prime}\{\lambda_{\nu^\prime}\}$, does not change the scaling of $I$ but
confines the eigenvalues to a finite interval, $0 \leq \tilde\lambda_\nu \leq
1$.  Power laws are most easily described on logarithmic scales.  Therefore, the
logarithmically integrated spectral density
\begin{equation}
  \label{eq:logint_ev_dens}
  \tilde{I}(\log \tilde\lambda) = 
  \log \left[ \frac{1}{N} \sum_{\nu=1}^N 
    \Theta( \log \tilde\lambda - \log \tilde\lambda_\nu) \right]
\end{equation}
is used such that a power-law target density $I^{\text{target}}(\lambda) \propto
\lambda^{\ds/2}$ appears as linear relation $\tilde{I}^{\text{target}}(\log
\tilde\lambda) = (\ds/2) \log \tilde\lambda$.  The spectral distance $\sdist$
to the evolution target is defined as
\begin{equation}
  \label{eq:sdist}
  \sdist( \tilde{I}, \tilde{I}^{\text{target}} ) = 
  \int_{\log \tilde{\lambda}_{\min}^\ast}^0 
  \left| \tilde{I}(\log \tilde\lambda) - \tilde{I}^{\text{target}}(\log \tilde\lambda) \right|^2
  \mathrm{d}\log \tilde\lambda \,.
\end{equation}
The lower integration boundary $\log \tilde{\lambda}_{\min}^\ast$ is chosen such
that $\tilde{I}^{\text{target}}(\log \tilde{\lambda}_{\min}^\ast) =
\log(N^{-1})$.  For the numerical calculations in this work, the base 10
logarithm was used as denoted in the figures.


\end{document}